\documentclass[11pt]{article}
\usepackage{graphicx,dcolumn,amsmath,latexsym,amssymb}
\usepackage{amsfonts,subfigure,caption}
\usepackage{color,bm}
\usepackage{placeins}

\def\Bmp#1{ \begin{minipage}{#1} }
\def\Bmpc#1{ \begin{minipage}[c]{#1} }
\def\Bmpt#1{ \begin{minipage}[t]{#1} }
\def\Bmpb#1{ \begin{minipage}[b]{#1} }
\def\Emp{ \end{minipage} }

\DeclareMathOperator{\csch}{csch} 

\def\CC{{\mathbb{C}}}

\def\RR{{\mathbb{R}}}

\def\ZZ{{\mathbb{Z}}}

\def\O{{\mathcal{O}}}

\def\bA{{\mathbf{A}}}

\def\tA{{\widetilde{\bA}}}
\def\tlambda{{\widetilde{\lambda}}}

\def\hzeta{{\widehat{\zeta}}}

\def\Dpartial#1#2{ {\frac{\partial #1}{\partial #2} }}

\def\Dpartialn#1#2#3{ {\frac{\partial^{#3} #1}{\partial #2^{#3}}} }

\begin{document}

\title{Stability of Confined Vortex Sheets}
\author{Bartosz Protas}
\date{Department of Mathematics and Statistics, \\ 
McMaster University, Hamilton, ON, Canada \\ 
Email: {\tt bprotas@mcmaster.ca} \\ \bigskip\bigskip
\today}

\maketitle

\begin{abstract}
  We propose a simple model for the evolution of an inviscid vortex
  sheet in a potential flow in a channel with parallel walls. This
  model is obtained by augmenting the Birkhoff-Rott equation with a
  potential field representing the effect of the solid boundaries.
  Analysis of the stability of equilibria corresponding to flat sheets
  demonstrates that in this new model the growth rates of the unstable
  modes remain unchanged as compared to the case with no confinement.
  Thus, in the presence of solid boundaries the equilibrium solution
  of the Birkhoff-Rott equation retains its extreme form of
  instability with the growth rates of the unstable modes increasing
  in proportion to their wavenumbers. This linear stability analysis
  is complemented with numerical computations performed for the
  nonlinear problem which show that confinement tends to accelerate
  the growth of instabilities in the nonlinear regime.
\end{abstract}

\section{Introduction}
\label{sec:intro}

Shear layers play an important role in fluid mechanics as they appear
in many flows of industrial and geophysical significance when boundary
layers separate from solid objects. A key property of shear layers is
that under typical conditions they are unstable and undergo the
Kelvin-Helmholtz instability as a result of which the vorticity from
the shear layer rolls up into big vortices. When occurring
recurrently, this phenomenon can in turn give rise to a turbulent
cascade. In this investigation we are interested in a simple inviscid
model of the Kelvin-Helmholtz instability occurring in a channel with
solid walls. In the context of more realistic flows this problem was
studied using an approach based on the theory of viscous potential
flows in \cite{funada_joseph_2001} where the effect of various problem
parameters on the growth rates of unstable modes was analyzed.
Analogous questions relevant for the stability of confined jets in
circular geometries in the presence of heat and mass transfer were
considered in \cite{aaa14}.

Inviscid vortex sheets, represented as one-dimensional (1D) curves
across which the tangential velocity component exhibits a
discontinuity and evolving under their own induction in a potential
flow, have been frequently invoked as a mathematical abstraction of
actual viscous shear layers \cite{fundam:saffman1}. In unbounded
domains they admit an elegant description in terms of the
Birkhoff-Rott equation.  This singular integro-differential equation
has a number of interesting properties --- in particular, its
equilibrium solution representing a flat (undeformed) vortex sheet is
highly unstable to small-wavelength perturbations, a fact that
underlies the ill-posedness of the Birkhoff-Rott model
\cite{vsheet:Moore}. As a result, computational studies involving the
Birkhoff-Rott equation typically require some regularization in order
to track its long-time evolution, usually in the form of the
well-known ``vortex-blob'' approach \cite{k86a} or using the more
recent Euler-alpha strategy \cite{hnp06a}.  A generic feature
characterizing the evolution of (regularized) inviscid vortex sheets
is roll-up producing localized vortex spirals \cite{k86b}.  There are
many interesting mathematical questions concerning various aspects of
the Birkhoff-Rott equation and we refer the reader to the collection
\cite{c89a} and monograph \cite{mb02} for further details on this
topic. {The problem of vortex sheet roll-up was recently
  revisited in \cite{DevoriaMohseni2018}.}  As was shown in
\cite{ProtasSakajo2018}, despite the severe instability of the
Birkhoff-Rott equation, the equilibrium corresponding to the flat
sheet can be efficiently stabilized using methods of modern control
theory. {We add that inviscid vortex sheets endowed with mass
  inertia and bending rigidity have been used to model flutter
  phenomena and results concerning the stability of a flapping flag in
  a channel were reported in \cite{Alben2015}.}

While the Birkhoff-Rott equation has been originally applied on
unbounded or laterally unbounded domains (i.e., domains periodic in
the streamwise direction and unbounded in the transverse direction),
in this study we consider vortex sheets confined to a {\em bounded}
domain with parallel walls representing a channel. By {considering}
the Birkhoff-Rott equation modified to account for the presence of
such solid boundaries, we derive a dynamical system describing the
evolution of a confined vortex sheet and demonstrate that,
interestingly, confinement does not change {the key features of
  the linear instability of the vortex sheet.} More specifically, in
the presence of solid boundaries the growth rates of the unstable
modes remain unchanged as compared to the original case with no
confinement. We note that this particular conclusion can also be
deduced as a special limiting case from Rayleigh's classical stability
analysis of parallel flows \cite{Rayleigh1894,Drazin2004book}.
{Numerical solution of the nonlinear problem indicates that
  confinement in fact enhances the growth of instabilities in the
  nonlinear regime.}  The structure of the paper is as follows: in the
next section we first recall the Birkhoff-Rott equation and show how
it can be modified to account for the effect of solid boundaries;
then, in Section \ref{sec:stab} we study the {linear} stability
of the equilibria in the new model, whereas {in Section
  \ref{sec:nonlinear} we present results obtained from the numerical
  solution of the nonlinear problem;} some final comments are deferred
to Section \ref{sec:final}.

\section{Inviscid Vortex Sheets}
\label{sec:sheets}

\begin{figure}
\centering
\includegraphics[width=0.6\textwidth]{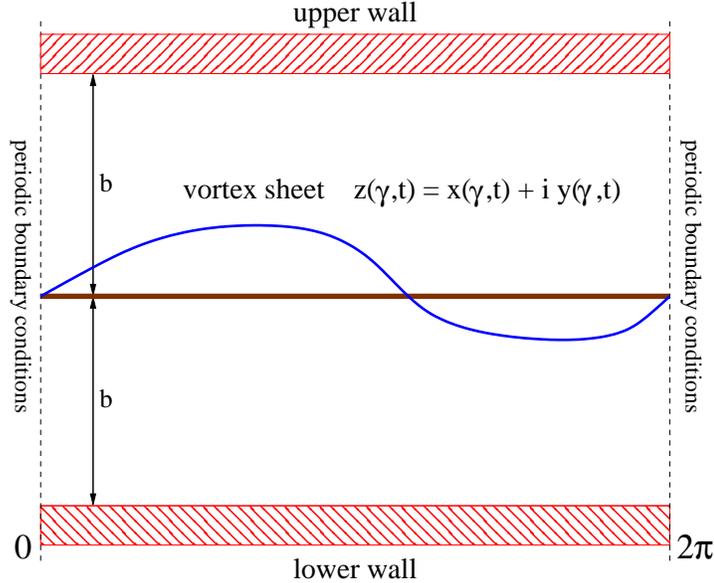}
\caption{Schematic representation of a confined vortex sheet. The
  thick brown line represents the equilibrium configuration
  $\tilde{z}(\gamma)$, $\gamma \in [0,2\pi]$.}
\label{fig:sheet}
\end{figure}

In this section we introduce a model for an inviscid vortex sheet
confined in a channel with two parallel walls located symmetrically
above and below the sheet in its equilibrium configuration, cf.~Figure
\ref{fig:sheet}. We assume that the flow is periodic, with period
$2\pi$, in the streamwise direction $x$. As is common in the study of
such problems \cite{fundam:saffman1}, we will extensively use the
complex representation of different quantities and will identify a
point $(x,y) \in \RR^2$ in the 2D space with $z = x + iy \in \CC$ in
the complex plane, where $i = \sqrt{-1}$ is the imaginary unit.  Let
then $z(\gamma,t)$ denote the position of a point (in the fixed frame
of reference) on the sheet which corresponds to the circulation
parameter $\gamma \in [0,2\pi]$ and some time $t$. The quantity
$\gamma$ represents a way of parameterizing the sheet and for sheets
of constant intensity is proportional to the arc-length of the curve.
Periodicity of the sheet then implies
\begin{equation}
z(\gamma+ 2 \pi, t)=z(\gamma,t)+2\pi, \quad \gamma \in [0,2\pi].
\label{eq:z}
\end{equation}
When the sheet is in a laterally unbounded domain, its evolution is
governed by the Birkhoff-Rott equation \cite{fundam:saffman1}
\begin{equation}
\Dpartial{z^\ast}{t}(\gamma,t) = V(z(\gamma,t))
:= \frac{1}{4\pi i}\mbox{pv} \int_0^{2\pi} 
\cot\left( \frac{z(\gamma,t)-z(\gamma^\prime,t) }{2} \right)\,d\gamma^\prime, 
\label{eq:BR}
\end{equation}
where $z^\ast$ denotes the complex conjugate of $z$, the integral on
the right-hand side (RHS) is understood in Cauchy's principal-value
sense and $V(z) = (u-iv)(z)$ represents the complex velocity at the
point $z$ with $u$ and $v$ the horizontal and vertical velocity
components (``:='' means ``equal to by definition'').

We now consider time evolution of a confined vortex sheet on a domain
$\Omega := \{ (x,y) \; : \; x \ \text{is $2\pi$-periodic}, \ -b < y < b
\}$, with straight boundaries $\partial\Omega$ located at $y = \pm b$
for some $b > 0$, cf.~Figure \ref{fig:sheet}. Distance $b$ will
serve as the main parameter in the problem. Since there is no flow
through these solid boundaries, the velocity in the flow must satisfy
the following conditions on its wall-normal component
\begin{equation}
\Im\left( V(x \pm i b ) \right) = 0.
\label{eq:BC}
\end{equation}
Since our model is inviscid, the tangential (slip) velocity component
on the walls $\Re\left( V(x \pm i b ) \right)$ need not vanish. In
order to satisfy condition \eqref{eq:BC}, the velocity field induced
by the vortex sheet, given by the RHS of the Birkhoff-Rott equation
\eqref{eq:BR}, must be augmented by including a suitable potential
velocity field $W(z) := \Dpartial{\phi}{x}(z) - i
\Dpartial{\phi}{y}(z)$ expressed in terms of a potential $\phi$. This
potential is constructed to cancel the normal component of the
velocity induced by the vortex sheet on the boundary $y = \pm b$, such
that the modified Birkhoff-Rott equation takes the following form
\begin{subequations}
\label{eq:BRb}
\begin{align}
\Dpartial{z^\ast}{t}(\gamma,t) & = V(z(\gamma,t)) + W(z(\gamma,t)) \nonumber \\
& = \frac{1}{4\pi i}\mbox{pv} \int_0^{2\pi} 
\cot\left( \frac{z(\gamma,t)-z(\gamma^\prime,t) }{2} \right)\,d\gamma^\prime \nonumber \\
& \phantom{=} + \Dpartial{\phi}{x}(z(\gamma,t)) - i \Dpartial{\phi}{y}(z(\gamma,t)), 
{\quad \gamma \in [0,2\pi], \ t > 0,}
\label{eq:BRb0} \\
\left(\Dpartialn{}{x}{2} + \Dpartialn{}{y}{2} \right) \phi &= 0 \qquad \text{in} \ \Omega, \label{eq:BRbLap} \\
\phi(x,y) & = \phi(x+2\pi,y), \qquad x \in (0,2\pi), \ -b < y < b, \label{eq:BRbper} \\
\Dpartial{\phi}{y}\bigg|_{y=\pm b} & = \Im\left[\frac{1}{4\pi i} \int_0^{2\pi} 
\cot\left( \frac{x \pm i b -z(\gamma^\prime,t) }{2} \right)\,d\gamma^\prime \right], \qquad x \in (0,2\pi), \label{eq:BRbb}
\end{align}
\end{subequations}
where relations \eqref{eq:BRbper}--\eqref{eq:BRbb} are the boundary
conditions for the Laplace equation \eqref{eq:BRbLap} defining the
potential $\phi$. We emphasize that the potential depends linearly on
the velocity induced by the vortex sheet on the solid boundaries $y =
\pm b$, cf.~\eqref{eq:BRbb}. Equation \eqref{eq:BR} and system
\eqref{eq:BRb} are complemented with a suitable initial condition
$z(\gamma,0) = z_0(\gamma)$, $\gamma \in [0,2\pi]$. As can be readily
verified, they admit an equilibrium solution (a fixed point)
$\tilde{z}(\gamma,t) := \gamma$ for $\gamma \in [0,2\pi]$, i.e.,
$\Dpartial{}{t}\tilde{z}^\ast(\gamma,t) = 0$, which corresponds to a
flat (undeformed) sheet, cf.~Figure \ref{fig:sheet}. The stability of
this equilibrium solution, and in particular how it depends on the
parameter $b$ defining confinement, is the main question addressed in
this paper.

We note that since the sheet intensity, which is assumed constant and
equal to unity, represents the difference between the tangential
velocity components on both sides of the sheet, the horizontal
velocity component in the flow is defined up to an arbitrary constant
$u_0$ which, however, does not affect the stability properties of the
sheet. To see this, we perform a change of coordinates to a moving
frame of reference $Z(t,\gamma) := z(t,\gamma) - u_0 t$, such that
equation \eqref{eq:BRb0} becomes $\Dpartial{Z^\ast}{t}(\gamma,t) =
V(Z(\gamma,t) + u_0 t) + W(Z(\gamma,t) + u_0 t) - u_0$. It is now
clear that the Jacobian of the RHS of this equation, and hence also
the stability properties of the sheet, do not depend on the constant
term $u_0$. Therefore, without loss of generality, below we will
assume that $u_0 = 0$.  The stability of confined equilibria described
by system \eqref{eq:BRb} is studied in the next section.

\section{Stability Analysis}
\label{sec:stab}

In this section we first review the stability properties of unconfined
sheet equilibria governed by the Birkhoff-Rott equation \eqref{eq:BR},
which are classical results \cite{ka79a,vsheet:SaOk96}, and then
proceed to analyze the stability of confined sheet equilibria
described by system \eqref{eq:BRb}. As a starting point, we perturb
the equilibrium state infinitesimally as
\begin{equation}
z({\gamma}, t)=x + \varepsilon \zeta({\gamma},t),
\label{eq:zpert}
\end{equation}
for some $0 < \varepsilon \ll 1$. Here, $\zeta({\gamma},t)$ is a perturbation
represented in the periodic setting, cf.~\eqref{eq:z}, as
\begin{equation}
\zeta({\gamma},t) =  \sum_{k=-\infty}^\infty \hzeta_k(t) \mbox{e}^{i k {\gamma}},
\label{eq:zeta}
\end{equation}
in which $\hzeta_k \in \CC$, $k \in \ZZ$, are the Fourier
coefficients.  Then, we obtain the linearized equation for the
evolution of the perturbation $\zeta({\gamma},t)$ as follows
\cite{vsheet:SaOk96}
\begin{eqnarray}
\frac{\partial \zeta^\ast}{\partial t}({\gamma},t) &=& V'[\zeta]({\gamma},t) := -\frac{1}{8 \pi i} \mbox{pv} \int_0^{2\pi} \frac{\zeta({\gamma},t)-\zeta({\gamma}^\prime,t)}{\sin^2\left( \frac{{\gamma}-{\gamma}^\prime}{2} \right)} \,d{\gamma}^\prime  \nonumber \\
 &=& \sum_{k=-\infty}^\infty \hzeta_k(t) \mbox{e}^{i k {\gamma}} \left[ -\frac{1}{8\pi i} \mbox{pv} \int_0^{2\pi} \frac{1-\mbox{e}^{-i k {\gamma}^\prime}}{\sin^2\left(\frac{{\gamma}^\prime}{2}\right)}\, d{\gamma}^\prime  \right] \nonumber \\
 &=& -\frac{1}{2i} \sum_{k=1}^\infty k \, \hzeta_k(t) \mbox{e}^{i k {\gamma}} -\frac{1}{2i} \sum_{k=1}^\infty  k \,\hzeta_{-k}(t) \mbox{e}^{-i k {\gamma}}, \label{eq:BRlin}
\end{eqnarray}
where $V'[\zeta]$ is a linear operator acting on $\zeta$ obtained as
the linearization of the RHS of equation \eqref{eq:BR} around the
equilibrium configuration.  Since
\begin{equation}
\frac{\partial \zeta^\ast}{\partial t} = \sum_{k=-\infty}^\infty \frac{\mbox{d} \hzeta^\ast_k}{\mbox{d}t} \mbox{e}^{-ik{\gamma}}
 = \sum_{k=1}^\infty \frac{\mbox{d} \hzeta_k^\ast}{\mbox{d}t} \mbox{e}^{-ik{\gamma}} + \sum_{k=1}^\infty \frac{\mbox{d} \hzeta_{-k}^\ast}{\mbox{d}t} \mbox{e}^{ik{\gamma}} + \frac{\mbox{d}\hzeta_0}{\mbox{d} t},
\label{eq:dzetadt0}
\end{equation}
equating coefficients of the Fourier components in \eqref{eq:BRlin}
and \eqref{eq:dzetadt0} corresponding to different wavenumbers $k$, we
obtain an infinite system of linear ordinary differential equations
(ODEs) for the evolution of the coefficients $\hzeta_k$ in a
block-diagonal form
\begin{subequations}
\label{eq:Leq}
\begin{align}
\frac{\mbox{d} \hzeta_0}{\mbox{d} t} &= 0, \label{eq:Leq0} \\
\frac{d}{dt}\begin{bmatrix} \hzeta_k \\ \hzeta_{-k}^\ast \end{bmatrix}
& = \frac{i\, k}{2}\begin{bmatrix} 0 & -1 \\ 1 & \phantom{-} 0 \end{bmatrix} 
\begin{bmatrix} \hzeta_k \\ \hzeta_{-k}^\ast \end{bmatrix}
=: \bA_k \begin{bmatrix} \hzeta_k \\ \hzeta_{-k}^\ast \end{bmatrix}, \qquad k = 1,2,\dots.
 \label{eq:Leqk}
\end{align}
\end{subequations}
corresponding to the original continuous integro-differential problem.
Since this latter system has a block-diagonal structure, key insights
about the infinite-dimensional problem can be obtained by analyzing
just a single block, independently from all other blocks.  The
eigenvalues corresponding to each diagonal block with matrix $\bA_k$,
$k \ge 1$, cf.~\eqref{eq:Leqk}, are $\lambda_k = \pm \frac{k}{2}$.
Thus, we see that at each wavenumber $k$ there is an unstable and
stable mode and the growth rate of the former is proportional to the
wavenumber $k$, such that small-scale perturbations always become more
unstable. This extreme form of instability underlies the ill-posedness
of the initial-value problem for the Birkhoff-Rott equation
\cite{vsheet:Moore}.

We now go on to analyze the stability of equilibria of confined vortex sheets
governed by system \eqref{eq:BRb} whose linearization takes the form
\begin{equation}
\Dpartial{\zeta^\ast}{t}({\gamma},t) = V'[\zeta]({\gamma},t) + W'[\zeta]({\gamma},t) 
\label{eq:lBRb}
\end{equation}
in which $W'[\zeta]({\gamma}) = \Dpartial{\phi'}{x}({\gamma},y=0) -i
\Dpartial{\phi'}{y}({\gamma},y=0)$ represents the linearization of the
potential velocity $W$ in \eqref{eq:BRb0} around the equilibrium. At
every instant of time $t$ the linearized potential $\phi'$ solves the
problem
\begin{subequations}
\label{eq:dphi}
\begin{alignat}{2}
\left(\Dpartialn{}{x}{2} + \Dpartialn{}{y}{2} \right) \phi' &= 0 & \qquad & \text{in} \ \Omega, \label{eq:dphi0} \\
\phi'(x,y) & = \phi'(x+2\pi,y), & \qquad & x \in (0,2\pi), \ -b < y < b, \label{eq:dphi1} \\
\Dpartial{\phi'}{y}\bigg|_{y=\pm b} & = \Im\left[ V'[\zeta](x \pm i b,t) \right], & \qquad & x \in (0,2\pi), \label{eq:dphi2}
\end{alignat}
\end{subequations}
where 
\begin{equation}
V'[\zeta](x \pm i b,t) = \frac{1}{8 \pi i} \int_0^{2\pi} \frac{\zeta({\gamma}',t) \, d{\gamma}'}
{ \sin^2\left(\frac{x \pm i b - {\gamma}'}{2} \right) }.
\label{eq:V'}
\end{equation}
Noting \eqref{eq:zeta} and the fact that $V'[\zeta]$ is linear in the
perturbation $\zeta$, the RHS in the boundary condition
\eqref{eq:dphi2} can be expressed as
\begin{align}
\Im\left[ V'[\zeta](x \pm i b,t) \right] &= \Im \left[\frac{1}{8 \pi i}  
\sum_{k=-\infty}^\infty \hzeta_k(t) \, I_k(x \pm i b) \right], \label{eq:dphi2b} \\
\text{where} \quad I_k(x \pm i b) & :=  \int_0^{2\pi} \frac{ e^{i k {\gamma}'} \, d{\gamma}'}
{ \sin^2\left(\frac{x \pm i b - {\gamma}'}{2} \right) }, \qquad k \in \ZZ,
\label{eq:Ik}
\end{align}
represents the linearized complex velocity induced by the Fourier
component of the perturbation $\zeta$ with wavenumber $k$,
cf.~\eqref{eq:zeta}, on the solid boundaries.

Our goal is to eliminate the linearized potential $\phi'$ such that
the term $W'[\zeta]$ in \eqref{eq:lBRb} can be expressed explicitly as
a function of the perturbation $\zeta$, which in turn will allow us to
assess the effect of the confinement on the stability of the
equilibrium $\tilde{z}$. This will be done in three simple steps
described in the following three sections.

\subsection{Evaluation of integrals $I_k$ in \eqref{eq:Ik}}
\label{sec:Ik}

The integrals $I_k$, $k \in \ZZ$, are defined in the classical
(Riemann) sense (i.e., they are nonsingular) and will be evaluated
using the calculus of residues. To this end, we map the integral
\eqref{eq:Ik} to the positively oriented unit circle $|z| = 1$ on the
complex plane $\CC$ using the change of variables $z = e^{i
  {\gamma}}$, such that $dz = i z \, d{\gamma}$ and
\eqref{eq:Ik} becomes
\begin{equation}
I_k(\xi) = 4 \, i \, \oint_{|z|=1} \frac{z^k \, dz}{e^{i\xi} - 2z + e^{-i\xi} z^2} 
= 4 \, i \, e^{i\xi} \, \oint_{|z|=1} \frac{z^k \, dz}{(z - e^{i\xi})^2}, 
\quad \text{where} \ \ \xi := x \pm i b.
\label{eq:Ikz}
\end{equation}
In order to evaluate the integral in \eqref{eq:Ikz} using the calculus
of residues, we decompose its integrand expression into partial
fractions, noting that $k$ can take both positive and negative values,
\begin{equation}
\frac{z^k}{(z - e^{i\xi})^2} = \frac{A_1}{z} + \dots + \frac{A_{|k|}}{z^k}  + 
\frac{B_1}{z - e^{i\xi}} + \frac{B_2}{(z - e^{i\xi})^2} + C_0 + C_1 z + \dots + C_{|k|-2} z^{|k|-2},
\label{eq:PF}
\end{equation}
where $A_1,\dots,A_{|k|},B_1,B_2,C_0.C_1,\dots, C_{|k|-2}\in \CC$,
such that
\begin{equation*}
\oint_{|z|=1} \frac{z^k \, dz}{(z - e^{i\xi})^2} = 2 \pi i \, \left[ A_1 + B_1 \right].
\end{equation*}
We note that $A_1 = \dots = A_{|k|} = 0$ if $k \ge 0$ and $C_0 = C_1 =
\dots = C_{|k|-2} = 0$ if $k \le 0$.  We then have the following four
cases depending on the signs of $k$ and $y = \Im(\xi) = \pm b$:
\begin{itemize}
\item $k \ge 0$ and $y = b > 0$ ($|e^{i \xi}| < 1$), such that $A_1 =
  0$, $B_1 = k e^{(k-1) i \xi}$ and
\begin{equation*}
\oint_{|z|=1} \frac{z^k \, dz}{(z - e^{i\xi})^2} = 2 \pi i \, \left[  k e^{(k-1) i \xi} \right],
\end{equation*}

\item $k \ge 0$ and $y = -b < 0$ ($|e^{i \xi}| > 1$), such that $A_1 =
  B_1 = 0$, and
\begin{equation*}
\oint_{|z|=1} \frac{z^k \, dz}{(z - e^{i\xi})^2} = 0,
\end{equation*}

\item $k < 0$ and $y = b > 0$ ($|e^{i \xi}| < 1$), such that $A_1 = -
  B_1 = - k e^{(k-1) i \xi}$, $A_1 + B_1 = 0$, and
\begin{equation*}
\oint_{|z|=1} \frac{z^k \, dz}{(z - e^{i\xi})^2} = 0,
\end{equation*}

\item $k < 0$ and $y = - b < 0$ ($|e^{i \xi}| > 1$), such that $A_1 =
  - k e^{(k-1) i \xi}$, $B_1 = 0$ and
\begin{equation*}
\oint_{|z|=1} \frac{z^k \, dz}{(z - e^{i\xi})^2} = - 2 \pi i \, \left[  k e^{(k-1) i \xi} \right].
\end{equation*}
\end{itemize}
Thus, we finally obtain
\begin{equation}
I_k(\xi) = \left\{
\begin{alignedat}{2}
- & 8 \pi k e^{i k \xi},& \quad & k \ge 0 \ \text{and} \ \Im(\xi) > 0 \\
+ & 8 \pi k e^{i k \xi},& \quad & k < 0 \ \text{and} \ \Im(\xi) < 0 \\
  & 0, & & \text{otherwise}
\end{alignedat} \right.,
\label{eq:Ikz2}
\end{equation}
a result which can be verified by approximating integrals
\eqref{eq:Ikz} numerically.

\subsection{Solution of System \eqref{eq:dphi} for Linearized Potential}
\label{sec:dphi}

Given the linearity of \eqref{eq:dphi0} and the form of the boundary
condition in \eqref{eq:dphi2}, which involves a superposition of
Fourier components with different wavenumbers $k$, we represent the
perturbation potential in the form of a series 
\begin{equation}
\phi'(x,y) = \sum_{k=0}^\infty \phi'_k(x,y ), \quad \text{where} \quad 
\phi'_k(x,y) := \frac{1}{2} ke^{-kb} \left[P_k(y) \, e^{i k x} + P_k^*(y) \, e^{- i k x} \right]
\label{eq:dphik}
\end{equation}
for some functions $P_k \; : \; [-b,b] \rightarrow \CC$,
$k=0,1,\dots$.  Using formulas \eqref{eq:Ikz2} and representation
\eqref{eq:dphik}, the boundary condition \eqref{eq:dphi2} becomes
equivalent to the following set of relations
\begin{subequations}
\label{eq:DphiDy}
\begin{align}
\Dpartial{\phi'_k}{y}\bigg|_{y = b} & = \Im\left[ i k \hzeta_k e^{- k b}e^{i k x} \right]
= \frac{k e^{-k b}}{2} \left( \hzeta_k  e^{i k x} +  \hzeta_k^*  e^{-i k x} \right), \\
\Dpartial{\phi'_k}{y}\bigg|_{y = - b} & = \Im\left[ i k \hzeta_{-k} e^{- k b}e^{- i k x}  \right]
= \frac{k e^{-k b}}{2} \left( \hzeta_{-k}  e^{-i k x} +  \hzeta_{-k}^*  e^{i k x} \right),
\quad k \ge 0,
\end{align}
\end{subequations}
such that system \eqref{eq:dphi} reduces to a family of 1D
boundary-value problems
\begin{subequations}
\label{eq:dPk}
\begin{align}
& \frac{d^2}{dy^2} P_k(y) - k^2 \, P_k(y) = 0, \qquad \text{in} \ (-b,b), \quad k = 0,1,2,\dots,  \\
& \frac{d}{dy} P_k(b) = \hzeta_k, \qquad \frac{d}{dy} P_k(-b) = \hzeta_{-k}^*. 
\end{align}
\end{subequations}
Their solutions are, noting that $\hzeta_0 \equiv 0$,
\begin{subequations}
\label{eq:Pk}
\begin{align}
P_0(y) &= 0,  \\
P_k(y) &= \left( \hzeta_k - \hzeta_{-k}^* \right) \frac{\cosh(k y)}{2 k \sinh(kb)}
+ \left( \hzeta_k + \hzeta_{-k}^* \right) \frac{\sinh(k y)}{2 k \cosh(kb)}, \qquad k \ge 1,
\end{align}
\end{subequations}
such that the perturbation potential finally becomes
\begin{equation}
\begin{aligned}
\phi'(x,y) = \sum_{k=1}^\infty \frac{k e^{-k b}}{2} & \left\{  \left[\left( \hzeta_k - \hzeta_{-k}^* \right) \frac{\cosh(k y)}{2 k \sinh(kb)} 
+ \left( \hzeta_k + \hzeta_{-k}^* \right) \frac{\sinh(k y)}{2 k \cosh(kb)}  \right] \, e^{i k x} \right. \\
& \hspace*{-0.15cm} + \left. \left[\left( \hzeta_k^* - \hzeta_{-k} \right) \frac{\cosh(k y)}{2 k \sinh(kb)} 
+ \left( \hzeta_k^* + \hzeta_{-k} \right) \frac{\sinh(k y)}{2 k \cosh(kb)}  \right] \, e^{-i k x}\right\}.
\end{aligned}
\label{eq:dphif}
\end{equation}

\subsection{Stability of Modified System \eqref{eq:BRb}}
\label{sec:BRb}

In order to evaluate the term $W'[\zeta]({\gamma},t)$ in \eqref{eq:lBRb}, we
need to compute the partial derivatives of the perturbation potential
\eqref{eq:dphif} at the location of the unperturbed sheet $\tilde{z}$,
i.e., at $y = 0$,
\begin{subequations}
\label{eq:DphiDx0}
\begin{align}
\Dpartial{\phi'}{x}\bigg|_{y = 0} & = \frac{1}{2} \sum_{k=1}^\infty k e^{-k b} \left[ i k P_k(0)  e^{i k x} - i k P_k^*(0)  e^{-i k x} \right], \\
\Dpartial{\phi'}{y}\bigg|_{y = 0} & = \frac{1}{2} \sum_{k=1}^\infty k e^{-k b} \left[ \frac{d}{dy}P_k(0)  e^{i k x} + \frac{d}{dy} P_k^*(0) e^{-i k x} \right].
\end{align}
\end{subequations}
{Using \eqref{eq:Pk} and recognizing that under the assumption of
  unit sheet strength at equilibrium we have $\gamma = x$, the
  following expression is obtained}
\begin{equation}
W'[\zeta]({\gamma},t) = \frac{i}{2} \sum_{k=1}^\infty \frac{k e^{-k b}}{\sinh(2kb)} \left[ \left( e^{-kb} \hzeta_k -  e^{kb} \hzeta_{-k}^* \right)e^{i k {\gamma}}  
-   \left( e^{kb} \hzeta_k^* -  e^{-kb} \hzeta_{-k} \right)e^{-i k {\gamma}} \right].
\label{eq:dW}
\end{equation}
Inserting this expression into \eqref{eq:lBRb} and rewriting this
equation in terms of Fourier components as was done earlier for the
original problem \eqref{eq:Leq}, we obtain an infinite system of ODEs
for the Fourier coefficients preserving the block-diagonal structure
of the original problem, with the block corresponding to the
wavenumber $k$ given by
\begin{subequations}
\label{Leqb}
\begin{align}
\frac{\mbox{d} \hzeta_0}{\mbox{d} t} &= 0, \label{Leqb0} \\
\frac{d}{dt}\begin{bmatrix} \hzeta_k \\ \hzeta_{-k}^\ast \end{bmatrix}
& = \frac{i\, k}{2} \begin{bmatrix} f_k(b) & -\left( 1 + g_k(b) \right) \\ 
\left( 1 + g_k(b) \right) &  - f_k(b) \end{bmatrix} 
\begin{bmatrix} \hzeta_k \\ \hzeta_{-k}^\ast \end{bmatrix} \nonumber \\
& = \frac{i\, k}{2} \begin{bmatrix} \csch(2kb) & -\coth(2kb) \\ 
\coth(2kb)  &  - \csch(2kb) \end{bmatrix} 
\begin{bmatrix} \hzeta_k \\ \hzeta_{-k}^\ast \end{bmatrix} 
=: \tA_k \begin{bmatrix} \hzeta_k \\ \hzeta_{-k}^\ast \end{bmatrix}, \qquad k = 1,2,\dots,
 \label{Leqbk}
\end{align}
\end{subequations}
where the expressions 
\begin{subequations}
\label{eq:fgk}
\begin{align}
f_k(b) & := \frac{1}{\sinh(2kb)}, \label{eq:fk} \\
g_k(b) & := \frac{e^{-2kb}}{\sinh(2kb)} \label{eq:gk}
\end{align}
\end{subequations}
represent ``corrections'' due to confinement effects.

\begin{figure}
\centering
\includegraphics[width=0.7\textwidth]{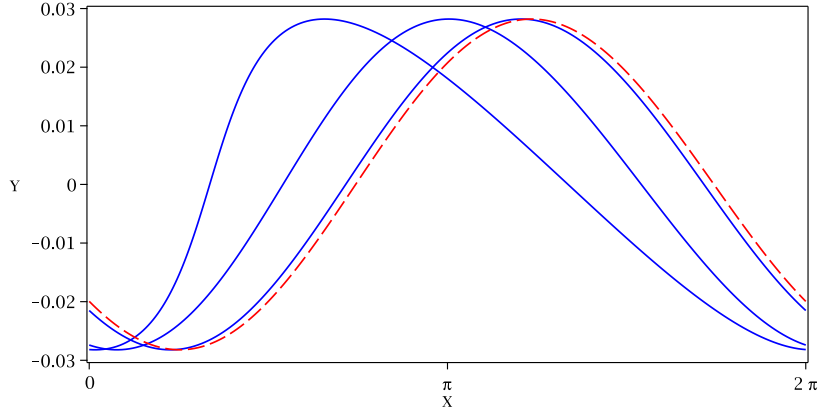}
\caption{Equilibria deformed as $\tilde{z} + \epsilon u_1$ with the
  unstable eigenvectors $u_1$ of (red dashed line) the unconfined
  problem \eqref{eq:Leq} and (blue solid lines) the confined problem
  \eqref{Leqb} with $b = 0.05, 0.25, 1.25$ (more deformed sheets
  correspond to smaller values of $b$). In all cases we have $k = 1$
  and $\epsilon = 0.05$, whereas the eigenvectors are normalized
  {such that $\int_0^{2\pi} \left[\Im(u_1(\gamma'))\right]^2 \,
    d\gamma' = 1$ which ensures the same mean-square transverse
    displacement in all cases, thereby making it easier to compare the
    deformations corresponding to different values of $b$.} The
  deformed equilibria are shifted by an appropriate distance in the
  horizontal direction so that their horizontal extent is $[0,2\pi]$
  in all cases.}
\label{fig:evec}
\end{figure}
Since when $k>0$ we have $\lim_{b \rightarrow 0} f_k(b), g_k(b) =
\infty$ and $\lim_{b \rightarrow \infty} f_k(b), g_k(b) = 0$, the
terms representing the effect of confinement in matrix $\tA_k$ vanish
when the distance $b$ increases such that, as expected, the unconfined
case is recovered in the limit $b \rightarrow \infty$. On the other
hand, these terms become dominant as the solid walls approach the
vortex sheet.  However, the eigenvalues of
matrix $\tA_k$ do not depend on the distance $b$.  They are given by
$\tlambda_k = \pm {\frac{k}{2}}$ and are identical to the
eigenvalues of matrix $\bA_k$, cf.~\eqref{eq:Leqk}, describing the
linearized evolution of unconfined vortex sheets. Thus, interestingly,
within the inviscid model considered here, confinement has no effect
on the growth rates of the unstable modes.

We now consider how confinement affects the form of unstable
eigenmodes and sheet equilibria deformed as $\tilde{z} + \epsilon u_1$
by the unstable modes $u_1$ with wavenumber $k=1$ are shown in Figure
\ref{fig:evec} for different values of parameter $b$.  Because the
perturbation amplitude $\epsilon = 0.05$ is relatively small, the
deformed equilibrium in the unconfined case resembles a {slanted sine
  wave (although the slant is imperceptibly small due to the vertical
  scale in Figure 2 being magnified with respect to the horizontal
  scale)}.  Since functions $f_k(b)$ and $g_k(b)$,
cf.~\eqref{eq:fk}--\eqref{eq:gk}, vanish very rapidly as $b$
increases, differences between the unstable modes in the unconfined
and confined cases are significant only for very small values of $b$,
of order $\O(10^{-1})$ or less, and are essentially imperceptible when
$b$ is $\O(1)$ or larger, which is the more relevant case from the
practical point of view. The effect of the confinement is to steepen
the profile of the unstable modes, {such that the eigenmodes with
  the same wavenumber $k$ become increasingly non-normal as $b$
  decreases. The influence of the confinement on the growth of
  instabilities in the nonlinear regime is explored using numerical
  computations in the next section.}

{
\section{Nonlinear Growth of Instabilities}
\label{sec:nonlinear}

In this section we solve system \eqref{eq:BRb} numerically for
different values of the parameter $b$ in order to characterize the
growth of instabilities beyond the linear regime discussed in Section
\ref{sec:stab}. System \eqref{eq:BRb} is discretized in space using
the collocation approach devised in \cite{vsheet:SaOk96}, but without
regularization, on a {mesh} consisting of $N = 128$ equispaced grid
points, see also \cite{ProtasSakajo2018}. The normal velocity
component on the channel boundaries is evaluated by approximating the
integral in \eqref{eq:BRbb} with the trapezoidal quadrature (which is
spectrally accurate for smooth functions defined on periodic domains).
After computing the discrete Fourier transform of this quantity, the
velocity components due to potential $\phi$ are determined
analytically as described in Section \ref{sec:stab},
cf.~\eqref{eq:dphik}--\eqref{eq:DphiDx0}, with the difference that
they are now evaluated at the actual sheet location $z(\gamma,t)$
rather than on the line $y=0$. Given the ill-posedness of the
Birkhoff-Rott equation resulting in its extreme sensitivity to
round-off errors, all calculations are performed with increased
arithmetic precision, typically using 64 significant digits, which is
done using {\tt Advanpix}, a Multiprecision Computing toolbox for
MATLAB \cite{Advanpix}.  The resulting ODE system is integrated in
time using a multiprecision version of the routine {\tt ode45}.  The
relative and absolute tolerance are both set to $10^{-8}$ which in the
course of extensive tests was found to ensure converged results.  We
note that with sufficient arithmetic precision there {is} no need to
use the spectral-filtering technique introduced by \cite{k86b} to
control the spurious growth of round-off errors.  }

{We use the same initial condition as originally employed in
  \cite{k86b}, i.e., $z_0(\gamma) = \gamma + 2 \pi \epsilon_0 (1-i)
  \sin\gamma$, $\gamma \in [0,2\pi]$, where $\epsilon_0 = 0.01$, which
  represents the equilibrium configuration perturbed with the unstable
  eigenmode of the unconfined problem with wavenumber $k=1$ (because
  of how our domain is defined, this initial condition is scaled by a
  factor of $2\pi$ as compared to \cite{k86b}). We emphasize that
  since in the absence of regularization the Birkhoff-Rott system
  \eqref{eq:BR} develops a curvature singularity in finite time
  \cite{vsheet:Moore,k86b,DevoriaMohseni2018}, our computations in the
  nonlinear regime can be carried out for relatively short times only
  when the solution remains well resolved.}

\begin{figure}
\begin{center}
\includegraphics[width=0.55\textwidth]{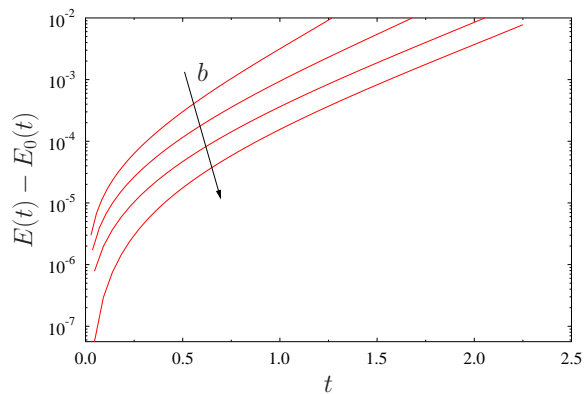}
\caption{{The difference $E(t) - E_0(t)$ between the perturbation
    energy in the nonlinear and linear evolution,
    cf.~\eqref{eq:E0}--\eqref{eq:E}, as a function of time $t$ for $b =
    0.25, 0.5, 1, 10$. The arrow indicates the trend with the increase
    of $b$.}}
\label{fig:Et}
\end{center}
\end{figure}

\begin{figure}
\begin{center}\hspace*{-0.5cm}
\mbox{\subfigure[]{\includegraphics[width=0.55\textwidth]{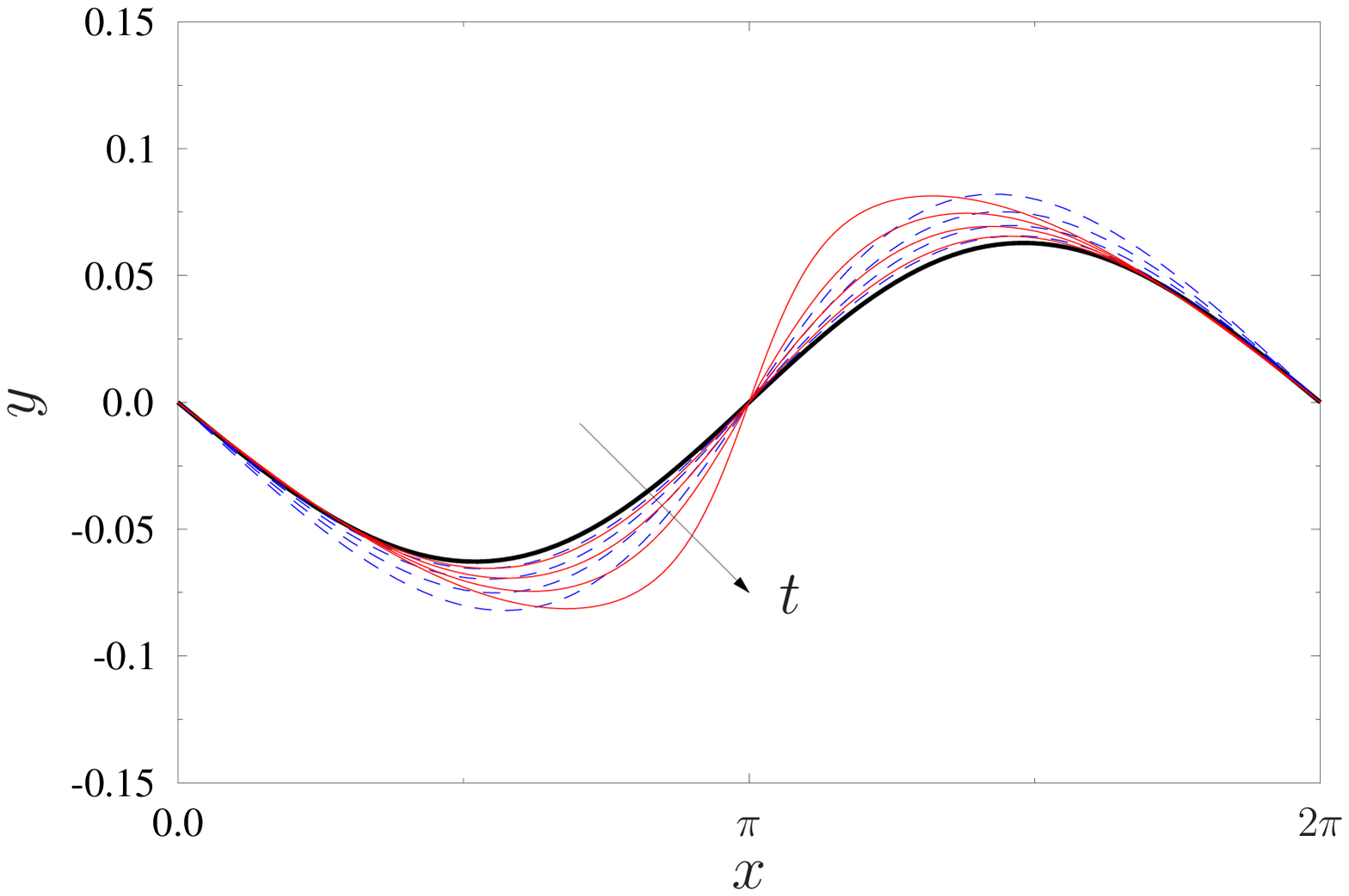}}
\subfigure[]{\includegraphics[width=0.55\textwidth]{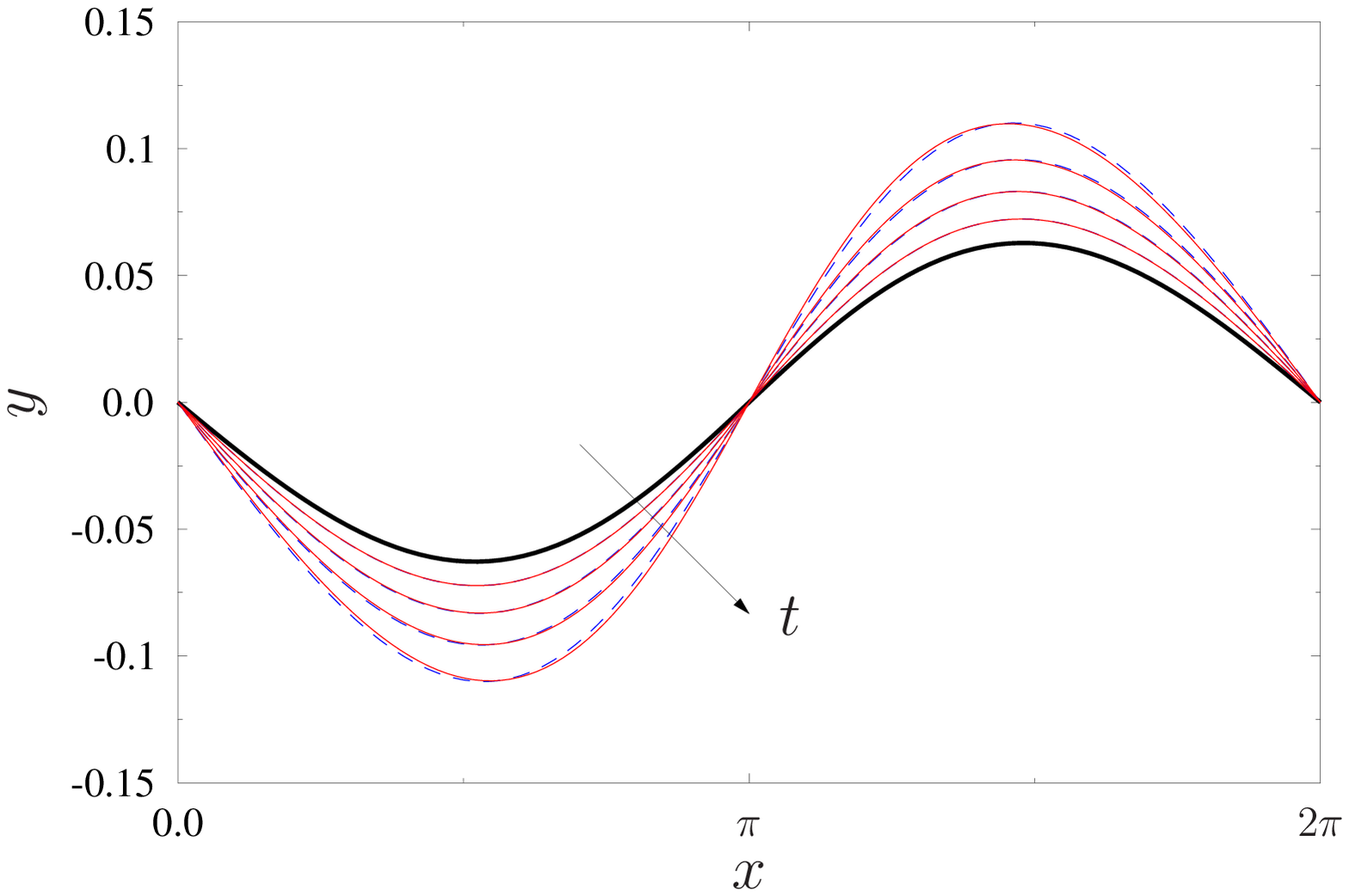}}}
\caption{{Solutions of (a) the confined problem with $b = 0.25$
    and (b) the unconfined problems at five equispaced time levels in
    the interval $[0,1.375]$. Dashed blue lines and solid red lines
    represent solutions of the nonlinear and linear problem
    \eqref{eq:BRb} and \eqref{eq:lBRb}--\eqref{eq:V'}, respectively,
    whereas the thick black line corresponds to the initial condition
    $z_0$. The arrows indicate the trend with the increase of time
    $t$.}  }
\label{fig:z}
\end{center}
\end{figure}

{In order to characterize the evolution of perturbations in the
  linear and nonlinear regime we define the following two quantities
\begin{subequations}
\label{eq:E0E}
\begin{align}
E_0(t) & := \epsilon_0^2 \int_0^{2\pi} | \zeta(\gamma,t)|^2 \, d\gamma, \label{eq:E0} \\
E(t) & := \int_0^{2\pi} | z(\gamma,t) - \gamma|^2 \, d\gamma. \label{eq:E}
\end{align}
\end{subequations}
We note that these quantities represent the sum of the ``energy''
associated with the transverse displacement of the sheet and its
longitudinal deformation, the latter of which is related to the
perturbation of the initially uniform circulation density of the
sheet.  Since in all cases quantities \eqref{eq:E0}--\eqref{eq:E}
evolve in a similar way, in Figure \ref{fig:Et} we show their
difference $E(t) - E_0(t)$ as a function of time $t$ for different
values of $b$. We see that as $b$ decreases the difference between the
nonlinear and linear evolution becomes more significant. The case
corresponding to $b=10$, which is the largest value of this parameter
we considered, is essentially indistinguishable from the unconfined
case. On the other hand, for smaller values of $b$ the curvature
singularity tends to occur earlier, so the nonlinear evolution can
only be computed for shorter times in these cases.}

{Solutions to the nonlinear and linear system \eqref{eq:BRb} and
  \eqref{eq:lBRb}--\eqref{eq:V'} obtained for $b=0.25$ are shown at
  different times in the physical space in Figure \ref{fig:z}(a). We
  note that in this case with significant confinement the perturbation
  amplitude $2\pi\cdot 0.01$ represents about $25\%$ of the channel
  half-width. For comparison, in Figure \ref{fig:z}(b) we show the
  corresponding solutions of the unconfined problems obtained at the
  same time levels. As is evident from these two figures, the primary
  effect of the confinement is to increase the ``steepness'' of the
  perturbed sheet while its transverse displacement is reduced with
  respect to the unconfined case. }

\section{Conclusions}
\label{sec:final}

We have proposed a simple model for the evolution of an inviscid
vortex sheet in a potential flow in a confined geometry representing a
channel with parallel walls. It is obtained by augmenting the standard
Birkhoff-Rott equation with a potential field representing the effect
of the solid boundaries. Next we considered the stability of
equilibria corresponding to flat sheets and demonstrated through
analytical computations that the presence of the solid boundaries does
not affect the block-diagonal structure of the modified Birkhoff-Rott
equation linearized around the equilibrium and, more importantly, the
growth rates of the unstable modes remain unchanged with respect to
the case with no confinement. Thus, in the presence of solid
boundaries the equilibrium solution of the Birkhoff-Rott equation
retains its extreme form of instability with the growth rates of the
unstable modes increasing in proportion to their wavenumbers. We note
that this finding is in fact also implied by {Rayleigh's classical
  analysis of the stability of parallel flows
  \cite{Rayleigh1894,Drazin2004book} as well as by} the results
reported in \cite{funada_joseph_2001}, although this observation {does
  not appear to have been explicitly stated before.}  As an
implication of this finding, we conclude that also in the presence of
confinement the Birkhoff-Rott equation cannot be integrated
numerically {past the Moore singularity \cite{vsheet:Moore} into the
  roll-up regime} unless some form of regularization is applied.
{Short-time numerical integration of nonlinear system
  \eqref{eq:BRb} carried out utilizing arbitrary precision arithmetics
  to control round-off errors indicates that confinement enhances the
  growth of perturbations in the nonlinear regime which is manifested
  by increased deformation of the perturbed sheet, cf.~Figures
  \ref{fig:Et} and \ref{fig:z}.}

The physical validity of the findings reported above is certainly
restricted by the assumptions inherent in our highly-idealized model,
most importantly, the assumption that the flow is irrotational away
from the vortex sheet. Due to boundary-layer effects, this assumption
is definitely not going to be satisfied for small values of the
distance $b$. As regards the dynamics of the vortex sheet itself,
viscous effects can be accounted for using the ``vortex-blob''
regularization approach \cite{k86a} or with the more recently proposed
Euler-alpha strategy \cite{hnp06a}. {The present study lays the
  groundwork for performing the stability analysis of such regularized
  models in the presence of confinement, with a view towards designing
  suitable control strategies based on these models (we emphasize here
  that earlier stability analyses of the vortex-sheet problem
  \cite{Rayleigh1894,Drazin2004book} do not in fact lend themselves to
  such extensions).}

\section*{Acknowledgments}

The author was supported through an NSERC (Canada) Discovery Grant. He
also acknowledges illuminating discussions about the Birkhoff-Rott
equation with Prof.~Takashi Sakajo. {Anonymous referees provided
  valuable comments on this work, in particular, regarding the
  connection to Rayleigh's classical analysis.}

\FloatBarrier 



\end{document}